\documentclass[aps,reprint,twocolumn,superscriptaddress,citeautoscript,showpacs,amsmath,amssymb,floatfix,prb]{revtex4-1}

\usepackage{amsmath}
\usepackage{amssymb}
\usepackage{graphicx}
\usepackage{times}
\usepackage{bm}
\usepackage{color}
\usepackage{gensymb}
\usepackage{dcolumn}

\newcolumntype{d}[1]{D{.}{.}{#1}}

\begin{document}

\title{High-resolution x-ray diffraction study of the heavy-fermion compound YbBiPt}

\author{B. G. Ueland}
\email{bgueland@ameslab.gov, bgueland@gmail.com}
\affiliation{Ames Laboratory, U.S. DOE, Iowa State University, Ames, Iowa 50011, USA}
\affiliation{Department of Physics and Astronomy, Iowa State University, Ames, Iowa 50011, USA}

\author{S. M. Saunders}
\affiliation{Ames Laboratory, U.S. DOE, Iowa State University, Ames, Iowa 50011, USA}
\affiliation{Department of Physics and Astronomy, Iowa State University, Ames, Iowa 50011, USA}

\author{S. L. Bud'ko}
\affiliation{Ames Laboratory, U.S. DOE, Iowa State University, Ames, Iowa 50011, USA}
\affiliation{Department of Physics and Astronomy, Iowa State University, Ames, Iowa 50011, USA}

\author{G. M. Schmiedeshoff}
\affiliation{Department of Physics, Occidental College, Los Angeles, California 90041, USA}

\author{P. C. Canfield}
\affiliation{Ames Laboratory, U.S. DOE, Iowa State University, Ames, Iowa 50011, USA}
\affiliation{Department of Physics and Astronomy, Iowa State University, Ames, Iowa 50011, USA}

\author{A. Kreyssig}
\affiliation{Ames Laboratory, U.S. DOE, Iowa State University, Ames, Iowa 50011, USA}
\affiliation{Department of Physics and Astronomy, Iowa State University, Ames, Iowa 50011, USA}

\author{A. I. Goldman}
\affiliation{Ames Laboratory, U.S. DOE, Iowa State University, Ames, Iowa 50011, USA}
\affiliation{Department of Physics and Astronomy, Iowa State University, Ames, Iowa 50011, USA}

\date{\today}

\begin{abstract}
YbBiPt is a heavy-fermion compound possessing significant short-range antiferromagnetic correlations below a temperature of $T^{\textrm{*}}=0.7$~K, fragile antiferromagnetic order below $T_{\textrm{N}}=0.4$~K, a Kondo temperature of $T_{\textrm{K}} \approx1$~K, and crystalline-electric-field splitting on the order of $E/k_{\textrm{B}}=1\,\textrm{--}\,10$~K.  Whereas the compound has a face-centered-cubic lattice at ambient temperature, certain experimental data, particularly those from studies aimed at determining its crystalline-electric-field scheme, suggest that the lattice distorts at lower temperature.  Here, we present results from high-resolution, high-energy x-ray diffraction experiments which show that, within our experimental resolution of $\approx6\,\textrm{--}\,10\times10^{-5}$~\AA, no structural phase transition occurs between $T=1.5$ and $50$~K.  In combination with results from dilatometry measurements, we further show that the compound's thermal expansion has a minimum at $\approx18$~K and a region of negative thermal expansion for  $9\alt T \alt18$~K.  Despite diffraction patterns taken at $1.6$~K  which indicate that the lattice is face-centered cubic and that the Yb resides on a crystallographic site with cubic point symmetry, we demonstrate that the linear thermal expansion may be modeled using crystalline-electric-field level schemes appropriate for Yb$^{3+}$ residing on a site with either cubic or less than cubic point symmetry.
\end{abstract}

\maketitle

\section{Introduction}
YbBiPt is a heavy-fermion compound which manifests an extraordinary Sommerfeld coefficient of $\approx8$~J/mol-K$^2$ and  spin-density-wave type antiferromagnetic (AFM) order below a N\'{e}el temperature of $T_{\textrm{N}}=0.4$~K  \cite{Canfield_1991,Fisk_1991,Movshovich_1994,Canfield_1994, Mun_2013, Ueland_2014}.  The low value of $T_{\textrm{N}}$ reflects the fact that the dominant magnetic energy scales are all small and comparable --- the Kondo temperature $T_{\textrm{K}} \approx1$~K \cite{Fisk_1991}, the Weiss temperature $\theta_{\textrm{W}} \approx-2$~K \cite{Mun_2013}, and the crystalline-electric-field (CEF) splitting is on the order of $1\,\textrm{--}\,10$~K \cite{Robinson_1995}.  As a consequence, the compound's electronic states are very responsive to small applied magnetic fields and pressures\cite{Mun_2013,Movshovich_1994b}, and recent neutron diffraction measurements have shown that the AFM order is quite fragile \cite{Ueland_2014}.   The magnetic phase diagram indicates that, as $T\rightarrow0$~K, AFM order persists up to a critical applied magnetic field of $\mu_{0}H_{\textrm{c}}\approx 0.4$~T, followed by a region of non-Fermi liquid behavior up to $\approx0.8$~T, above which Fermi-liquid behavior occurs up to at least $6$~T \cite{Mun_2013}.  It has been proposed that a magnetic-field-induced quantum-critical point occurs at $\mu_{0}H_{\textrm{c}}$ \cite{Mun_2013}.  

YbBiPt crystallizes in the face-centered-cubic half-Heusler structure (space group $F\overline{4}3m$) with a room temperature lattice parameter of $a=6.5953(1)$~\AA \cite{Robinson_1994}.  The Yb ions are located at the $4d$ Wyckoff position, forming a face-centered-cubic magnetic sublattice, and the Bi and Pt are located at the $4c$ and $4a$ Wyckoff positions, respectively \cite{Robinson_1994}.  All three sites possess $\overline{4}3m$ tetrahedral (i.e.\,cubic) point-group symmetry.  Neutron scattering experiments have shown that the ambient field magnetic order is characterized by an AFM propagation vector of $\bm{\tau}=(\frac{1}{2},~\frac{1}{2},~\frac{1}{2})$ with the magnetic moments directed parallel to $\bm{\tau}$ \cite{Ueland_2014}.  Surprisingly, the scattering at the magnetic Bragg positions was found to consist of two coincident peaks: a broad peak corresponding to a magnetic correlation length of $\xi_{\rm{b}} \approx 20$~{\AA}~and a narrower peak corresponding to AFM correlations extending over $\xi_{\rm{n}}\approx80$~\AA.  The broad peak appears upon cooling through $T^{\textrm{*}}=0.7$~K, while the narrower peak appears below $T_{\textrm{N}}$.  The total integrated intensity under both peaks corresponds to a magnetic moment of $\approx0.8~\mu_{\textrm{B}}$. However, the ratio of the integrated intensity of the broad peak to that of the narrower peak is $\approx12:1$, and previous experiments have estimated that the static ordered moment is $0.1$ to $0.25~\mu_{\textrm{B}}$ \cite{Robinson_1994,Amato_1992}.

Many questions regarding the nature of the low-temperature magnetism and the proposed quantum-critical point remain to be answered.  For example,  several previous studies have suggested that a lattice distortion occurs at $T\approx6$~K which results in a lowering of the cubic symmetry \cite{Robinson_1995,Martins_1995,Robinson_1999}.  Specifically, inelastic neutron scattering data suggest that the $J=7/2$ ground-state magnetic multiplet of the Yb is split by the CEF further than allowed for by cubic point symmetry \cite{Robinson_1995}, and specific heat data show a feature at $6$~K that cannot be described only by a simple Schottky-term \cite{Robinson_1999}.  In addition, electron spin resonance experiments on Er-doped YbBiPt suggest that slight distortions away from cubic symmetry occur in the vicinity of the rare-earth sites \cite{Martins_1995}, and measurements of the thermal expansion have shown that the thermal expansion coefficient changes sign close to $6$~K \cite{Mun_2013}.

\begin{figure*}
\centering
\includegraphics[width=1.0\linewidth]{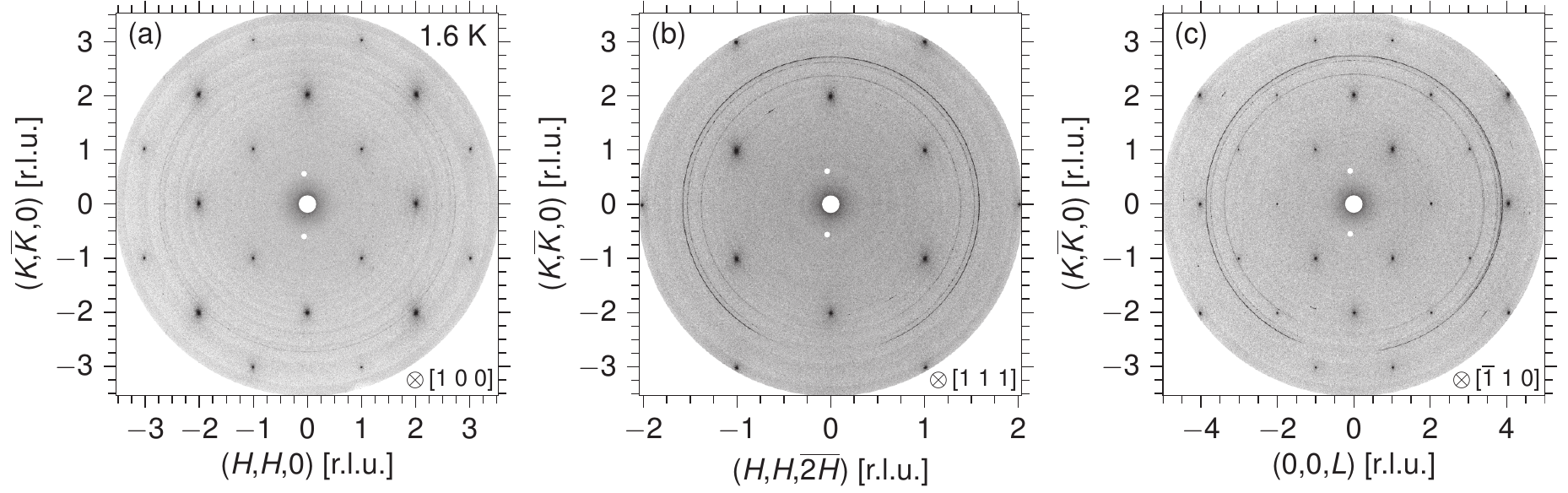}
\caption{Diffraction patterns taken at $T=1.6$~K using the MAR345 image plate with the x-ray beam parallel to the $[0~0~1]$~(a), $[1~1~1]$~(b), and $[\overline{1}~1~0]$~(c) crystal axes.  The center of the detector and two spurious points located above and below the center due to ghost images from a previously measured sample have been masked by white circles.  r.l.u. = reciprocal lattice units.}
\label{Fig1}
\end{figure*}

In this paper, we present results from high-resolution, high-energy x-ray diffraction experiments on single-crystal samples of YbBiPt.  Our data show that, within the experimental resolution, no bulk distortion of the lattice or the formation of a superstructure occur down to $T=1.5$~K.  In combination with data from dilatometry measurements, we further show that a minimum in the thermal expansion occurs at $\approx18$~K, followed by a region of negative thermal expansion for  $9\alt T \alt18$~K, and that both features may be modeled using either the CEF level scheme appropriate for cubic symmetry or a CEF level scheme appropriate for lower than cubic point symmetry.  We suggest that further measurements are necessary, in particular, inelastic neutron scattering experiments on single-crystal samples, in order to elucidate the CEF level scheme, and we discuss the importance of understanding the CEF level scheme in relation to the compound's complex low-temperature magnetism.

\section{Experiment}
High-quality single crystals of YbBiPt were grown using a Bi flux, as described previously \cite{Canfield_1991, Mun_2013}.  A sample approximately $1.5$~mm thick with smooth surfaces was chosen for the measurements, and excess flux was carefully removed from the surfaces prior to the experiments.  High-energy x-ray diffraction measurements were made at station $6$-ID-D at the Advanced Photon Source using an x-ray wavelength of $\lambda=0.09407$~\AA\ and a beam size of $60\times60~\mu\textrm{m}$.  The sample was cooled down to $T=1.5$~K using a He closed-cycle cryostat with a Joule-Thompson stage.  Two Be domes were placed over the sample and evacuated, and a small amount of He gas was subsequently added to the inner dome to facilitate thermal equilibrium.  A cylindrical aluminized-Kapton heat shield also surrounded the sample and inner Be dome.  Using \textsc{mucal} \cite{mucal} to determine the absorption coefficients of YbBiPt's constituents for $\lambda=0.09407$~\AA, and using the thickness of the sample and the beam size given above, we calculated that the transmission of the sample was $71$\%.  We therefore did not expect any significant heating of the sample by the x-ray beam.  We also did not observe any rise in temperature upon exposing the sample to the beam at $T=1.5$~K, and believe that the sample was in thermal equilibrium during the measurements.

The cryostat was mounted to the sample stage of a $6$-circle diffractometer, and a MAR$345$ image plate and Pixirad-$1$ area detector were used to measure the diffracted x-rays transmitted through the sample. The MAR$345$ image plate was positioned with its center aligned to the beamstop and was set back $2.411$~m  from the sample (as determined from measurement of a Si standard from the National Institute of Standards and Technology) to record a diffraction pattern spanning a scattering angle of $|2\theta|\lesssim4.1\degree$.  The detector was operated with a pixel size of $100\times100~\mu\textrm{m}^{2}$, which resulted in an angular resolution of $\Delta2\theta\approx0.0024\degree$.  The value for the resolution corresponds to changes in lattice plane spacings ($d$-spacings) of $1.3\times10^{-2}$\AA~for $2\theta=1\degree$ and $7.7\times10^{-4}$\AA~for $2\theta=4.1\degree$.  Data were taken by recording an image while tilting the sample along two rocking angles.

The Pixirad detector is comprised of a hexagonal array of pixels with a spacing of $60~\mu$m.  The detector was oriented such that it covered a $1.075\degree$~range in $2\theta$ across its horizontal axis, and its center was initially aligned to the direct beam.  Hence, the horizontal axis of the detector was lying approximately along the direction of the scattering vector \textbf{Q}.  Diffraction data for various Bragg peaks were recorded by rocking the sample around its vertical axis in step sizes ranging from $0.5\,\textrm{--}\,1\times10^{-3}$\degree.  A frame was recorded for each rocking step and divided by the corresponding number of monitor counts.  The frames were added together to produce a $2$-D image of a Bragg peak.

\section{Results}

Figure \ref{Fig1} shows diffraction patterns recorded at $T=1.6$~K using the MAR345 image plate for the incident beam along the three characteristic directions of cubic symmetry.  For space group $F\overline{4}3m$, there are no special reflection conditions for the sites occupied by Yb, Bi, and Pt. Therefore, the space-group symmetry does not exclude each element from contributing to each Bragg peak, and the general reflection conditions are such that all $(H~K~L)$ must be either even or odd.  The rings in the patterns are due to the Be domes attached to the displex and residual Bi flux on the sample.  Using the integrated intensities and structure factors for the the Bi $(1~1~0)$ and the YbBiPt $(2~2~0)$ Bragg peaks, we estimate that the residual Bi accounts for less than $3\%$ of the volume of the sample illuminated by the beam.  The plane-group symmetries observable in Figs.\,\ref{Fig1} (a)\,--\,(c) are consistent with the $F\overline{4}3m$ space group, namely: $p4mm$ for Fig.~\ref{Fig1}(a), $p31m$ for Fig.~\ref{Fig1}(b), and $c1m1$ for Fig.~\ref{Fig1}(c)\cite{ITC_2006}.    The angles between the Bragg peaks and the horizontal axis passing through the origin were examined for each figure and were found to be consistent with the values for a cubic lattice within an uncertainty of $\approx\pm0.01\degree$.  There is also no observable broadening or splitting of the diffraction peaks that would suggest a distortion away from cubic symmetry.  Finally, no additional Bragg peaks were observed, which rules out the formation of a superstructure or breaking of the face-centered-cubic symmetry.   The dynamic range of our measurements was $2.3\times10^{4}$:1 and was limited by the thickness and absorption of the sample.  Mo filters were used to attenuate the direct beam, and the transmission of the filters was $0.0313$ for Fig.~\ref{Fig1}(a), and $0.1769$ for Figs.~\ref{Fig1}(b) and \ref{Fig1}(c).

\begin{figure}
\centering
\includegraphics[width=1.0\linewidth]{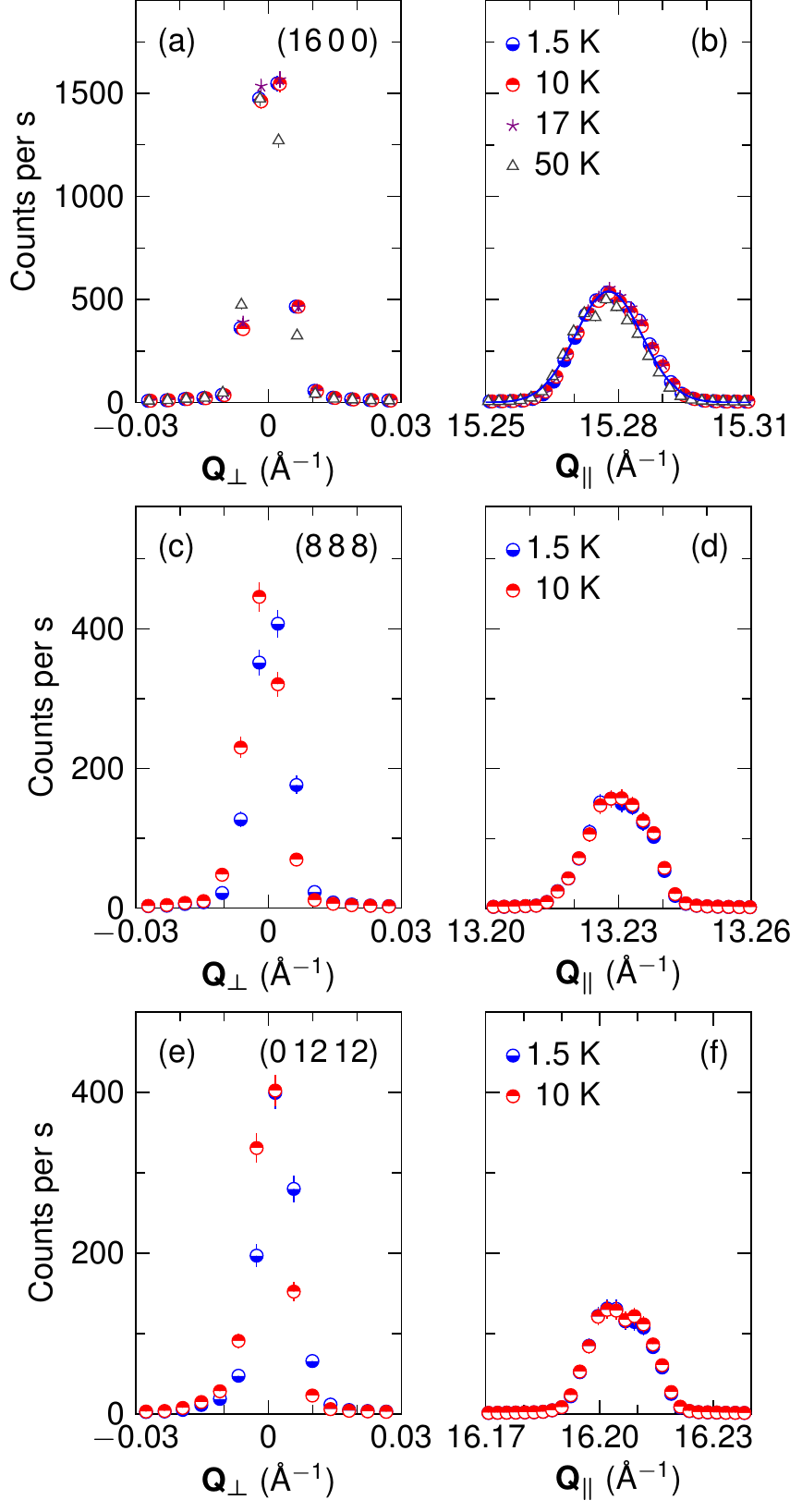}
\caption{Cuts through the $(16~0~0)$~[(a) and (b)], $(8~8~8)$~[(c) and (d)] and $(0~12~12)$~[(e) and (f)] Bragg peaks taken either perpendicular to (\textbf{Q}$_{\bm{\perp}}$) or parallel to (\textbf{Q}$_{\bm{\parallel}}$) the scattering vector \textbf{Q}, at various temperatures.   The construction of the curves is described in the text, and the scattering intensities have been multiplied by an average monitor rate of $6658$~counts/s.  The line in (b) shows a fit of  a Gaussian line shape to the $T=1.5$~K data.}
\label{Fig2}
\end{figure}

Figure \ref{Fig2} shows cuts through the $(16~0~0)$, $(8~8~8)$, and $(0~12~12)$ Bragg peaks, at various temperatures, from data taken with the Pixirad area detector.  The curves were constructed by summing the detector's pixels along either the horizontal or vertical direction to generate cuts along directions perpendicular to (\textbf{Q}$_{\bm{\perp}}$) and parallel to (\textbf{Q}$_{\bm{\parallel}}$) the scattering vector \textbf{Q}, respectively.  By rotating $2\theta$ with the sample aligned at the $(16~0~0)$ Bragg-peak position ($2\theta=13.134\degree$), we determined that the spacing between pixels along the horizontal direction corresponds to $2.1 \times 10^{-3}\,\degree\, (2\theta)$ per pixel.  Using this value, we determined the sample to detector distance and also the spacing between pixels along the vertical direction, which is $3.6 \times 10^{-3}\,\degree\, (2\theta)$ per pixel.  These values correspond to changes in $Q$ of $\Delta Q_{\perp}=4.2\times 10^{-3}$~\AA$^{-1}$ and $\Delta Q_{\parallel}=2.4\times 10^{-3}$~\AA$^{-1}$.  $\Delta Q_{\parallel}=2.4\times 10^{-3}$~\AA$^{-1}$ corresponds to a change in spacing between lattice planes of $d=6 \times 10^{-5}$~\AA, which may be compared to the precision for $a$ of $1\times10^{-4}$~\AA{} reported for previous neutron scattering experiments \cite{Robinson_1994}.  

Since the size of the incident x-ray beam was comparable to the spacing between pixels and the thickness of the illuminated region of the sample was $\approx 1.5$~mm, the shapes of the Bragg peaks are primarily governed by the sample's absorption and slight variations in the thickness of the sample as it is rotated.  This interpretation is supported by the fact that the diffraction patterns taken with the MAR345 image plate showed single peaks at each expected Bragg position at $1.6$~K. We fit the peaks in the \textbf{Q}$_{\bm{\parallel}}$ cuts to a Gaussian line shape, which only approximates the shape of the peak, but is sufficient for determining the center and estimating the full width at half maximum (FWHM).  The line in Fig.~\ref{Fig2}(b) shows the fit to the $(16~0~0)$ Bragg peak at $1.5$~K.  Its center is at $15.2774(1)$\AA$^{-1}$.  At $1.5$~K, the FWHMs of the Bragg peaks are $0.0089(1)$, $0.0112(2)$, and $0.0102(2)$~\AA$^{-1}$~for Figs.\,\ref{Fig2}(a), \ref{Fig2}(c), and \ref{Fig2}(e), respectively, and  $0.0170(2)$, $0.0156(3)$, and $0.0162(2)$~\AA$^{-1}$ for Figs.\,\ref{Fig2}(b), \ref{Fig2}(d), and \ref{Fig2}(f), respectively.  The FWHMs and peaks shapes did not change over the temperature ranges measured.  The change in the center position of the $(16~0~0)$ peak with increasing temperature is discussed below.

\begin{figure}
\centering
\includegraphics[width=1.0\linewidth]{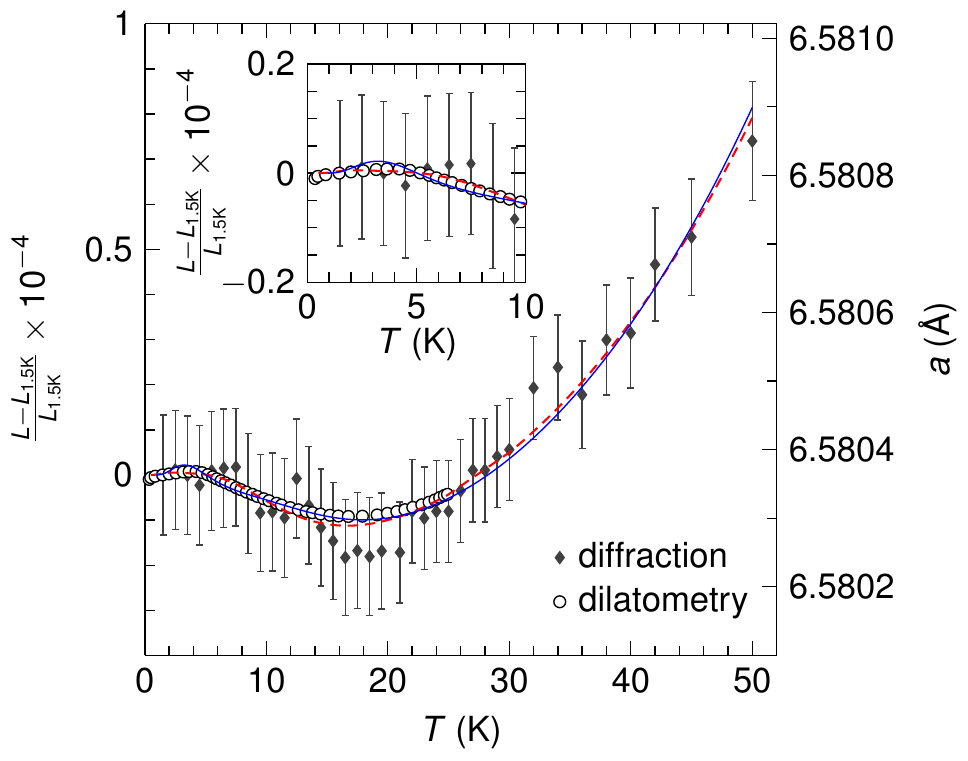}
\caption{Temperature dependence of the linear thermal expansion $\alpha\Delta T = \frac{L-L_{0}}{L_{0}}$ (left axis) with $L_{0}$ taken at $T=1.5$~K, and the lattice parameter $a$ (right axis) determined by fitting cuts of the  ($16~0~0$) Bragg peak along \textbf{Q}$_{\bm{\parallel}}$ to Gaussian line shapes (diamonds).   The circles show linear thermal expansion data taken with a capacitive dilatometer between $0.35$ and $25$~K.  The errorbars indicate the uncertainty in the diffraction data, while the uncertainty is within the symbol size for the dilatometry data.  The lines are simultaneous fits to the x-ray and dilatometry data, as described in the text.  The dashed red line corresponds to the doublet-quartet-doublet CEF level scheme predicted for the Yb$^{3+}$ site possessing cubic point symmetry.  The solid blue line corresponds to a CEF scheme consisting of $4$ doublets, which would correspond to the Yb$^{3+}$ site having lower than cubic point symmetry.  The inset shows a blow-up of the low temperature region.}
\label{Fig3}
\end{figure}

Measurements of the $(16~0~0)$ Bragg peak were made upon warming up to $T=50$~K.  The centers determined from fits to the cuts made along \textbf{Q}$_{\bm{\parallel}}$ to a Gaussian line shape were used to determine the temperature dependence of $a$ and the linear thermal expansion $\alpha\Delta T = \frac{L-L_{0}}{L_{0}}$, where $L_{0}$ corresponds to the measurement made at $1.5$~K.   $\alpha$ is the linear thermal expansion coefficient.  These data are shown by the diamonds in Fig.~\ref{Fig3}.  $a$ is constant (within the uncertainty) from $1.5$~K up to $\approx9$~K, then decreases with increasing temperature (which corresponds to a region of negative thermal expansion), reaches a minimum at $\approx18$~K, and then increases with increasing temperature up to $50$~K.  The circles show data from linear thermal expansion measurements \cite{Mun_2013} made along the $[1~0~0]$ crystalline direction between $0.35$ and $25$~K using a capacitive dilatometer \cite{Schmiedeshoff_2006}.  Here, $L_{0}$ also has been taken at $1.5$~K, and the data likewise show a minimum at $\approx18$~K followed by a region of negative thermal expansion at lower temperature.  One notable difference with the diffraction data is that the trend in the dilatometry data suggests that a weak maximum may occur at $\approx4$~K.  The lines in Fig.~\ref{Fig3} are fits to models described below.

\section{Discussion}
\subsection{Symmetry of the lattice at low temperature}
The diffraction patterns in Fig.~\ref{Fig1} are consistent with the lattice being described by a face-centered-cubic space group, which is the symmetry possessed by the lattice at $T=300$~K \cite{Robinson_1994}.  There are also no indications of the formation of a superstructure, such as the occurrence of additional Bragg peaks upon cooling.  In addition to the data shown in Fig.~\ref{Fig1}, data were taken using different amounts of attenuation, and include patterns with over-illuminated Bragg peaks.  No peaks indicative of a superstructure or of lowering of the face-centered-cubic symmetry were found in such over-illuminated patterns either.  Data taken with the Pixirad detector also show no evidence for a temperature at which splitting or anomalous broadening of the Bragg peaks occurs. Nevertheless, we have made two thorough analyses of the possible structures consistent with the observed low-temperature diffraction patterns, paying particular attention to the point-symmetry of the Yb site at low temperature.

First, an examination of  the Wyckoff positions for all of the face-centered-cubic space groups listed in Ref.~\onlinecite{ITC_2006} shows that the minimum multiplicity of the positions for such groups is either $4$ or $6$.  For those face-centered-cubic groups containing positions with a multiplicity of $4$, which is the number of Yb ions in a unit cell, the positions all possess cubic point symmetry.  Hence, based on the diffraction patterns, which indicate that the lattice is face-centered cubic at $T=1.6$~K, the Yb ions are located at positions with cubic point symmetry. 

Next, since the signature of a structural phase transition in experimental data for thermodynamic quantities may be weak, we also consider the consequences of a second-order structural phase transition occurring at some temperature between $T=300$ and $1.5$~K.   This analysis is performed in accordance with Landau's theory for a second-order phase transition;  in particular, if a second-order structural phase transition occurs, then the structures on either sides of the transition should be related through a group-subgroup relation \cite{ITC_2006b}.  The pertinent subgroups for  $F\overline{4}3m$ are the cubic groups $F23$, and $P\overline{4}3m$, the tetragonal group $I\overline{4}m2$, and the trigonal group $R3m$.  Since the symmetry of the diffraction patterns in Fig.~\ref{Fig1} corresponds to cubic symmetry, we rule out that the lattice is described best by a tetragonal or trigonal space group at $1.6$~K.  The peaks in the diffraction pattern fit the reflection conditions for a face-centered cell, namely, $h+k$, $h+l$, and $k+l=2n$ or $h$, $k$, $l$ all even or all odd.  This rules out the $P\overline{4}3m$ subgroup.  The remaining subgroup, $F23$, may not be ruled out.  However, following the argument given above, the $4$ Yb ions would still reside on a site with cubic point symmetry, since $F23$ is a face-centered-cubic space group.  Finally, the isomorphic subgroups $F\overline{4}3m$ possessing enlarged unit cells can be ruled out due to the lack of additional Bragg peaks that would correspond to such enlarged unit cells.

Summarizing, our analyses of the diffraction patterns show that YbBiPt has a face-centered-cubic lattice down to at least $T=1.5$~K, and that the Yb are located at a site with cubic point symmetry.  However, since our diffraction data are sensitive to the average, or global, symmetry of the compound, they may average over localized, uncorrelated lattice distortions that could potentially lower the cubic point symmetry of the Yb site.  In the next section, we describe how the point-symmetry of the Yb site relates to the energy levels formed by the CEF, and how the CEF level scheme affects the thermal expansion.

\subsection{Thermal expansion and the crystalline-electric-field scheme}
The degeneracy of the ground-state magnetic multiplet of the Yb ions should be at least partially lifted by the CEF formed by their neighbors.  For YbBiPt, the Yb site nominally possesses cubic point symmetry, and analysis of high-temperature magnetization data shows that the effective moment per Yb is $p_{\textrm{eff}}= 4.3~\mu_{\textrm{B}}$ \cite{Mun_2013}, which is close to the value of $4.5~\mu_{\textrm{B}}$ expected for Yb$^{3+}$ ($J=7/2$).  Using these facts, the point-charge model predicts that the cubic CEF should split the $J=7/2$ magnetic multiplet of the Yb$^{3+}$ into $\Gamma_{6}$ and $\Gamma_{7}$ doublets and a $\Gamma_{8}$ quartet \cite{Lea_1962,Dieke_1968}.  Next, specific heat data show that an entropy of $S=R\ln2$ is recovered by \mbox{$T\approx1$~K}, which indicates that the CEF ground state is likely a doublet \cite{Fisk_1991}, and data from specific heat and electron spin resonance experiments point to the first and second excited levels occurring at $E\approx0.5$ to $1.1$~meV and $E\agt7.6$~meV, respectively \cite{Mun_2013, Pagliuso_1999}.  Finally, from magnetic susceptibility data, the ground state was determined to likely be a $\Gamma_{7}$ doublet \cite{Fisk_1991}, and, from inelastic neutron scattering data, the first excited level was predicted to be a $\Gamma_{8}$ quartet \cite{Robinson_1993}.

On the other hand, inelastic neutron scattering experiments found a scattering peak at $E=5.7$~meV, which was identified as a CEF level transition to an excited doublet, and two inelastic peaks centered at $\approx1$~and $2$~meV, which become resolved at  $T\alt10$~K and lie on top of two broad quasielastic signals \cite{Robinson_1995}.  Experiments have not yet detected any dispersion associated with the  peaks at $\approx1$~and $2$~meV, which suggests that they may correspond to CEF level transitions.  As for the broad quasielastic signals, they are also seen above $10$~K, and it has been proposed that they arise from transitions occurring within the $\Gamma_{7}$ doublet and $\Gamma_{8}$ quartet \cite{Robinson_1995}.  Broad quasielastic scattering in other heavy-fermion compounds has been associated with the hybridization between the localized $4f$ electrons and itinerant charges and has been related to the value of $T_{\textrm{K}}$ \cite{Stewart_1984, Walter_1987}.   

If the two inelastic peaks at $E\approx1$ and $2$~meV correspond to CEF level transitions, then the point symmetry of the Yb$^{3+}$ sites would be lower than cubic, since for the cubic CEF level scheme only the $\Gamma_{7}\leftrightarrow\Gamma_{8}$ and $\Gamma_{8}\leftrightarrow\Gamma_{6}$ transitions have nonzero transition probabilities \cite{Birgeneau_1972}.  Based on the inelastic neutron scattering data, two scenarios for the CEF level scheme were proposed  \cite{Robinson_1995}: (1) a $\Gamma_{7}$ ground-state doublet lying close to an excited-state $\Gamma_{8}$ quartet, followed by a $\Gamma_{6}$ doublet located $\approx6$~meV above the nearly degenerate $\Gamma_{7}$ and $\Gamma_{8}$ levels; or ($2$) a ground-state doublet followed by $3$ excited doublets located at $\approx1$, $2$, and $6$~meV.  The latter scenario corresponds to the case of the Yb site possessing lower than cubic point symmetry.

The effect of the CEF level scheme on the thermal expansion may be modeled, and, here, we follow the derivations given in Refs.~\onlinecite{Ott_1976} and \onlinecite{Ott_1977}.  First,
the partition function for an ion with CEF levels at energies $E_{i}$ is:
\begin{equation}
Z=\displaystyle\sum_{i}g_{i}e^{\frac{-E_{i}}{k_{\textrm{B}}T}},
\label{Eq1}
\end{equation}
where $k_{\textrm{B}}$ is the Boltzmann constant, $i$ labels each CEF energy level, and $g_{i}$ is the degeneracy of each level.  The free energy density for $N$ non-interacting ions in a volume $V$ is $F=-k_{B}TN\ln Z$, and may be used to solve for the volume thermal expansion, which is given by:
\begin{equation}
\beta=-\kappa\frac{\partial^{2}F}{\partial V \partial T},
\label{Eq2}
\end{equation}
Here, $p$ is the applied pressure and $\kappa$ is the isothermal compressibility.  For a cubic crystal, $\beta=3\alpha$.  Next, by defining a Gr\"{u}neisen parameter $\gamma_{i}$ for a CEF level with an energy $E_{i}$ as: 
\begin{equation}
\gamma_{i}=-\frac{\partial\ln E_{i}}{\partial\ln V},
\label{Eq3}
\end{equation}
one can show that:
\begin{equation}
\beta=\frac{\kappa N}{k_{\textrm{B}}T^{2}}[\langle E^{2}\gamma\rangle-\langle E\gamma\rangle\langle E\rangle],
\label{Eq4}
\end{equation}
where the brackets denote the thermodynamic average.  As an example, thermal expansion data for TmSb show a minimum at $T=8$~K, which a fit using Eq.~\ref{Eq4} has shown is due to a CEF level transition with $E/k_{\textrm{B}}=25$~K \cite{Ott_1976}.  

The lines in Fig.~\ref{Fig3} show simultaneous least-square fits to both the x-ray diffraction and dilatometry data for $T\ge0.5$~K using the sum of Eq.~\ref{Eq4} and an $AT^{3}$ term.  The latter term accounts for the phonon contribution to the thermal expansion and $A$ is a constant.  The dashed red line corresponds to the doublet-quartet-doublet level scheme expected for a CEF with cubic symmetry acting on the Yb$^{3+}$ ions within YbBiPt.  For this fit, $\gamma_{i}$, $A$, and the energy of the excited doublet were allowed to vary, and the location of the quartet was taken as $E=0.52$~meV, which corresponds to the value obtained from electron spin resonance experiments on Y$_{0.9}$Yb$_{0.1}$BiPt \cite{Pagliuso_1999}.  The solid blue line corresponds to the CEF level scheme given in Ref.~\onlinecite{Robinson_1995}, which is appropriate for a CEF with lower than cubic symmetry acting on the Yb$^{3+}$ ions.  It consists of $4$ doublets located at $0$, $1$, $2$, and $6$~meV.  For this fit, only $\gamma_{i}$ and $A$ were allowed to vary.  Both of the fits appear adequate, reproducing the minimum at $\approx18$~K and the region of negative thermal expansion, however,the solid blue line shows a more pronounced maximum at $\approx4$~K.  Nevertheless, since $T_{\textrm{K}}$ and $\theta_{\textrm{W}}$ are on the order of $1$~K, and $T_{N}=0.4$~K and $T^{\textrm{*}}=0.7$~K, magnetic correlations may obfuscate the effect of the CEF level scheme on $\alpha\Delta T$ at $4$~K.

\begin{table}
 \caption{Parameters from the fits to the linear thermal expansion data using either the CEF level scheme expected for cubic symmetry or the CEF level scheme consisting of $4$ doublets.  The values \mbox{$N=1.4038\times10^{28}$~Yb/m$^{3}$} and $\kappa=1.0101\times10^{-11}$~m$^{2}$/N were used.  The value of $A$ is $7.0(2)\times10^{-10}$~$\textrm{K}^{-3}$ for the cubic CEF level scheme and $7.4(2)\times10^{-10}$~$\textrm{K}^{-3}$ for the CEF level scheme consisting of $4$ doublets.  \label{Tab1}}
 \begin{ruledtabular}
 \begin{tabular}{cd{0.3}d{0.5}|cd{0.2}r}
\multicolumn{3}{c|}{Cubic}&\multicolumn{3}{c}{$4$ doublets}\\
\hline
\multicolumn{1}{c}{Level}  & \multicolumn{1}{c}{$E$~(meV)} & \multicolumn{1}{c|}{$\gamma_{i}$}& \multicolumn{1}{c}{Level} & \multicolumn{1}{c}{$E$~(meV)} & \multicolumn{1}{c}{$\gamma_{i}$}\\
\hline
$\Gamma_{7}$ & 0 & \multicolumn{1}{c|}{---} & doublet & 0 &  \multicolumn{1}{c}{---} \\
$\Gamma_{8}$ & 0.52 & 0.1(2)& doublet & 1.0 & $1.5(6)$\\
$\Gamma_{6}$ & 4.3(2) & -14.8(7)& doublet & 2.0 &$-8(2)$\\
&&& doublet & 6.0 & $-14.1(7)$\\
 \end{tabular}
 \end{ruledtabular}
 \end{table}

Table \ref{Tab1} reports the parameters determined from the fits.  To the best of our knowledge, the compressibility of YbBiPt has not been reported, so we have used the relation $\kappa=1/B_{0}$ and a bulk modulus of $B_{0}=99$~GPa, which was calculated for LaBiPt  \cite{Oguchi_2001}.  The values for $\gamma_{i}$ differ, which, contrary to the usual invocation of the Gr\"{u}neisen parameter of  $\gamma=\frac{\beta}{\kappa C}$, suggests that the specific heat $C$ and $\beta$ are not proportional to each other over the whole temperature range of the experiment.  On the other hand, the large error in  $\gamma_{i}$ for the $\Gamma_{8}$ level of the cubic CEF level scheme and for the second excited doublet of the 4-doublet level scheme illustrates the shortcomings of our fits.  Specifically,  $\gamma_{i}$ and the energy of the CEF levels are correlated, and either fitting both simultaneously or having only weak features in the linear thermal expansion leads to large uncertainty in the fitted values.  Ideally, the energy of each CEF level should first be accurately determined through other methods (e.g. through specific heat, electron spin resonance, or inelastic neutron scattering measurements).  However, as mentioned above, the multiple competing  low-energy interactions in YbBiPt have made determining the CEF level scheme problematic.
 
While the fits to the data in Fig.~\ref{Fig3} cannot distinguish between the presence of a cubic or non-cubic CEF, it is apparent from our diffraction data that no splitting of the Bragg peaks occurs and that no peaks corresponding to a superstructure form upon cooling. As noted above, this means that the global symmetry of the compound remains face-centered-cubic down to $T=1.5$~K, at least within our experimental resolution.  Nevertheless, our diffraction data do not rule out the possibility of localized distortions of the crystal lattice that would lower the cubic point symmetry of the Yb sites.  In fact, $^{170}$Yb M\"{o}ssbauer spectroscopy measurements on YbBiPd and YbSbPb, which both have cubic lattices similar to that for YbBiPt, have shown that the local point symmetry of the Yb site is lowered away from cubic symmetry \cite{LeBras_1995}, and results from similar measurements on a polycrystalline sample of YbBiPt have been interpreted as indicating that only $\approx85$\% of the Yb ions are located at a site with cubic point symmetry \cite{LeBras_1994}.

 Our results suggest that new measurements, in particular inelastic neutron scattering measurements, on single-crystal samples are necessary to solve the CEF level-scheme for the following reasons:  ($1$) The peaks in the previously reported inelastic neutron scattering data located at $E\approx1$ and $2$~meV apparently showed no dispersion \cite{Robinson_1995}, however, the measurements were performed on polycrystalline samples, which means that any weak dispersion may not have been discernible.  ($2$) The previously reported low-temperature specific heat data for single-crystal samples show broad features above $T_{\textrm{N}}$ \cite{Mun_2013}, which makes a quantitative determination of the CEF-level scheme difficult.  ($3$) The previous electron spin resonance measurements were made on a small single-crystal sample of Y$_{0.9}$Yb$_{0.1}$BiPt \cite{Pagliuso_1999}, and the effects of spin-spin correlations on the CEF level scheme may not have been fully realized.  In addition, further measurements employing probes more sensitive to local distortions of the lattice, such as x-ray spectroscopy (in particular extended x-ray absorption fine structure), atomic-pair-distribution-function analysis, and M\"{o}ssbauer spectroscopy, should be made on single-crystal samples.

\section{Conclusion}
Using high-energy x-ray diffraction, we have shown that the global symmetry of YbBiPt's lattice remains face-centered cubic between  $T=1.5\,\textrm{--}\,50$~K, within the resolution of our experiments of $\approx6\,\textrm{--}\,10\times10^{-5}$~\AA, and that we find no evidence for the formation of a superstructure.  By considering the possible space groups consistent with our diffraction patterns taken at $1.6$~K, we have demonstrated that the patterns imply that the Yb ions reside on a site with cubic point symmetry.   In addition, we have shown that the linear thermal expansion possesses a minimum at $\approx18$~K and a region of negative thermal expansion for $9\alt T \alt18$~K, and that the data may be satisfactorily modeled using either a CEF level scheme appropriate for cubic symmetry or a CEF level scheme appropriate for lower than cubic symmetry.  We suggest that new inelastic neutron scattering measurements on single-crystal samples are necessary to fully solve the CEF level scheme, and that further measurements sensitive to localized distortions of the crystal lattice are necessary.  Finally, we note that while inelastic neutron scattering is particularly sensitive to magnetic dipole transitions, interactions involving higher-order multipole terms may be important to the low-temperature magnetism and the heavy-fermion type behavior of YbBiPt \cite{Santini_2009,Kuramoto_2009}.  While we do not address the issue of higher-order multipole ordering or transitions here, our results that the global symmetry of the lattice and that the site-symmetry of the Yb site remains cubic down to $1.5$~K may offer some constraints for models considered in future works.  In light of the multiple competing low-energy interactions present in YbBiPt, which are reflected in the broad features in the specific heat and the broad quasielastic scattering, as well as the significant AFM correlations evident below 0.7~K, determining the CEF level scheme appears to be vital to understanding the compound's complex magnetism.

\begin{acknowledgments}
We are grateful to D. S. Robinson for support during the x-ray experiments.  Work at the Ames Laboratory was supported by the Department of Energy, Basic Energy Sciences, Division of Materials Sciences \& Engineering, under Contract No. DE-AC02-07CH11358.  Work at Occidental College was supported by the National Science Foundation under DMR-1408598.  This research used resources of the Advanced Photon Source, a U.S. Department of Energy (DOE) Office of Science User Facility operated for the DOE Office of Science by Argonne National Laboratory under Contract No. DE-AC02-06CH11357.  
\end{acknowledgments}

\end{document}